\documentclass[twocolumn,prb,superscriptaddress,nofootinbib]{revtex4-1}
\usepackage[english]{babel}
\usepackage[ansinew]{inputenc}
\usepackage{boldline,multirow,xcolor,colortbl}
\usepackage{times}
\usepackage{graphicx}
\usepackage{graphics}
\usepackage{amsmath}
\usepackage{amsfonts}
\usepackage{amssymb}
\usepackage{amsbsy}
\usepackage{dsfont}
\usepackage{epstopdf}
\usepackage{makeidx}
\usepackage{subfigure}
\usepackage{color} 
\usepackage{pgf}
\usepackage{bm}
\usepackage{tikz} 
\usepackage[normalem]{ulem}
\usepackage[symbol]{footmisc}
\newcommand{\be}{\begin{equation}}
\newcommand{\ee}{\end{equation}}
\newcommand{\bea}{\begin{eqnarray}}
\newcommand{\eea}{\end{eqnarray}}

\setcitestyle{open={[},close={]}}
\begin{document}

\title{Nonlinear dynamics of topological Dirac fermions in 2D spin-orbit coupled materials}

%
\author{Rajesh K. Malla}
\affiliation{Theoretical Division, Los Alamos National Laboratory, MS B262, Los Alamos, New Mexico 87545, USA}
\affiliation{Center for Nonlinear Studies, MS B258, Los Alamos National Laboratory, Los Alamos, New Mexico 87545, USA}
\author{Wilton J. M. Kort-Kamp$^*$}
\affiliation{Theoretical Division, Los Alamos National Laboratory, MS B262, Los Alamos, New Mexico 87545, USA}
\begin{abstract}
The graphene family materials are two-dimensional staggered monolayers with a gapped energy band structure due to intrinsic spin-orbit coupling. The mass gaps in these materials can be manipulated on-demand via biasing with a static electric field, an off-resonance circularly polarized laser, or an exchange interaction field, allowing the monolayer to be driven through a multitude of topological phase transitions. We investigate the dynamics of spin-orbit coupled graphene family materials  to unveil topological phase transition fingerprints embedded in the nonlinear regime and show how these signatures manifest in the nonlinear Kerr effect and in third-harmonic generation processes.  We show that the resonant nonlinear spectral response of topological fermions can be traced to specific Dirac cones in these materials, enabling characterization of topological invariants in any phase by detecting the cross-polarized component of the electromagnetic field. By shedding light on the unique processes involved in harmonic generation via topological phenomena our findings open an encouraging path towards the development of novel nonlinear systems based on two-dimensional semiconductors of the graphene family.
\end{abstract}
\maketitle

\begin{figure}
\includegraphics[width=1.0\linewidth]{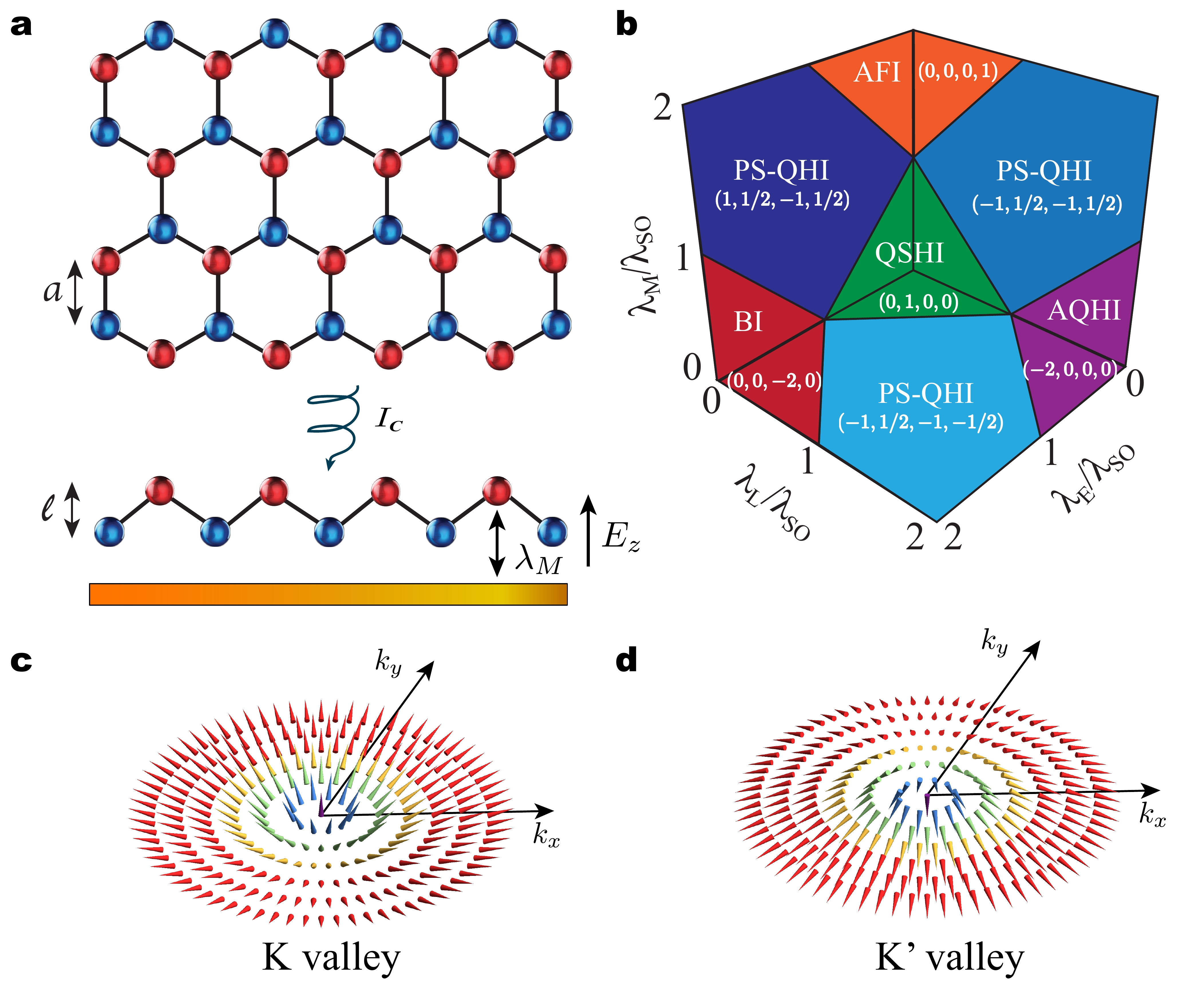}
\caption{{\bf Topological phases in the graphene family.} (a) Schematic representation of a graphene family monolayer under the influence of a static electric field ($E_z$), an off-resonance circularly polarized laser with intensity $I_C$, and the staggered exchange field $\lambda_M$. The in-plane lattice constant is $a$ and the buckling distance between sites $A$  and $B$ is $2l$. (b) 3D topological phase diagram and corresponding Chern numbers (${\cal C}$, ${\cal C}_s$, ${\cal C}_{\eta}$,  ${\cal C}_{s\eta}$) in spin-orbit coupled graphene family materials. The momentum dependent meron structures for a given spin are shown in (c) and (d) for valleys $K$ and (d) $K'$, respectively. }
\label{Fig1}
\end{figure}

 Graphene is the typical go-to material to investigate the optoelectronic response of two-dimensional (2D) systems  \cite{Neto2009, Peres2010} because of its extraordinary electron mobility \cite{Graphenereview}, tunable linear electronic conductivity \cite{Tunable2011}, and potential for strong-light matter interactions at sub-wavelength scales \cite{Grapheneplasmons}. Although it supports quantum Hall states in the presence of strong magnetic fields \cite{QHall1,QHall2, QHall3}, the intrinsically weak spin-orbit coupling \cite{WeakSOC} severely limits graphene's suitability to study topological phase transitions \cite{TI1,TI2} in low-dimensional materials.  The search for alternative  topologically non-trivial 2D structures with characteristics similar to graphene recently ended with the synthesis of silicene \cite{Silicene},  germanene \cite{Germanene},  stanene \cite{Stanene}, and plumbene \cite{Plumbene}.  The newer members of the graphene family materials (GFM) have a buckled honeycomb lattice with silicon, germanium, tin, or lead atoms occupying the lattice sites  \cite{Gomez2016, Molle:2017aa, Mannix:2017aa}. Unlike graphene, the strong spin-orbit coupling in these systems results in a gapped bulk energy band-structure and protected one-way edge states characteristic of topological insulators. Strikingly, they can be driven through a variety of phase transitions (Fig. \ref{Fig1}a,b) by actively controlling  the mass gap via external interactions \cite{EzawaJapanese}, which could enable unprecedented all-in-one material multi-optoelectronic functionalities with potential applications to spintronics and valleytronics.

 Various recent works have investigated ultrafast and nonlinear effects in topologically trivial 2D materials \cite{Gnonlinear1, Mikhailov2008, Kerr, Gnonlinear2, Transition, Exp2, Exp3, Exp4, Exp5, Sipe2014, Sipe2015, Pedersen2016, Mikhailov2016, HHG1,  Lysne2020, Linear1} and in topological systems \cite{HHG4, HHG5, HHG6, Chacon2020, RSeq}. Unravelling the interplay between topology and nonlinear effects in spin-orbit coupled monolayer semiconductors of the graphene family is a natural step at the materials science forefront, which could aid in developing next-generation technologies that meet the urgent demands for, among others, higher performance radio-frequency modulators, optically gated transistors, and practical spintronic-based devices. Nevertheless, studies on the optoelectronic properties of these materials have largely focused in their linear response\cite{Tunablebandgap1,Tunablebandgap2, WiltonPRB, WiltonPRL, WiltonNonlocal, Circular3, Wiltonnature, Tabert2014, Farias2018, Wu2018, Wu2020}, which include investigations of signatures of the tunable band gap \cite{Tunablebandgap1, Circular3}, spatial dispersion \cite{Tabert2014, WiltonNonlocal}, the interplay between the quantum Hall effect \cite{Tunablebandgap2} and photo-induced topology \cite{WiltonPRB}, as well as of topological phase transitions in quantum forces \cite{Wiltonnature, Farias2018}, spin-orbit photonic interactions \cite{WiltonPRL},  and light beam shifts \cite{Wu2018, Wu2020}.    The crossroads between nonlinear dynamics and topology in the extended graphene family, potentially allowing access to topological phase transition signatures, material symmetries, selection rules, and relaxation mechanisms otherwise screened by spurious effects in the linear response, remains uncharted.

We explore the interplay between topological Dirac fermions in the GFM and optical fields  beyond the linear regime. We develop a comprehensive and unified description of their nonlinear dynamical conductivity tensor, including non-trivial  topological  phases due to  an  externally applied electrostatic field,  a  circularly  polarized  laser, and exchange interaction.  Our results show a rich structure of nonlinear optoelectronic responses across the phase diagram, and we demonstrate that the third-order optoelectronic conductivity contribution to the fundamental and third-harmonics encodes fingerprints of topological phase transitions. We show that characterization of the Chern number is possible near bandgap resonances via polarization sensitive detection of the cross-polarized component of the field harmonics. Our work sets the cornerstones for investigating topological phase transitions beyond the linear response in the family of graphene-like elemental 2D materials.
\\

\hspace{-12pt}{\bf Results}

\hspace{-12pt}{\bf Topological properties of the graphene family - } 
 Let us consider the following Dirac-like Hamiltonian describing staggered GFM in the low-energy regime \cite{EzawaJapanese},
\begin{equation}
\label{Hamiltonian}
{\hat H}_s^{\eta}= {\boldsymbol{\tau}}\cdot {\bf d}^{\eta}_s= \hbar v_F(\eta k_x \tau_x +k_y \tau_y) + \Delta_s^{\eta} \tau_z,
\end{equation}
where $ {\bf d}^{\eta}_s = \{\eta \hbar v_F k_x, \hbar v_F k_y, \Delta_s^{\eta}\}$ describes a meron structure in momentum space (Fig. \ref{Fig1}c,d), $\hat{\textbf{p}}= \hbar(k_x,k_y)$ is the particle momentum, $\tau_i$ are the Pauli matrices, $\eta, s = \pm 1$ are valley and spin indexes, $v_F$ is the Fermi velocity, and the mass term $\Delta_s^{\eta}=\eta s \lambda_{SO}-\lambda_{E}-\eta \lambda_{L}+s\lambda_{M}$ corresponds to half energy band gap. Here, $\lambda_{SO}$ represents the intrinsic spin-orbit coupling energy and the last three terms in $\Delta_s^\eta$ account for interactions with external fields, which allow for tailoring the Dirac mass gap. Indeed, the second term $\lambda_E$  corresponds to the potential difference between sub-lattices in the buckled structure in the presence of a static electric field $E_z$ applied perpendicularly to the plane of the monolayer \cite{Tunablebandgap1}. The third component $\lambda_L$ describes the anomalous quantum Hall effect and arises due to the coupling between an off-resonant circularly polarized laser and the GFM \cite{Circular2,Tunablebandgap2}. The final term $\lambda_M$ depicts the staggered antiferromagnetic exchange interaction \cite{Antiferromagnetic}. For simplicity, we have neglected Rashba couplings due to their small magnitude compared to the other contributions. It is worth mentioning that the Hamiltonian in Eq. \eqref{Hamiltonian} applies more generally to other 2D systems, including  antiferromagnetic manganese chalcogenophosphates (MnPX$_3$, X = S, Se) \cite{Li} and perovskites \cite{Liang}. The formalism developed here is also valid for these systems.

 Within the Dirac picture the electronic properties of the GFM can be fully characterized through a set of four topological invariants \cite{EzawaJapanese}, namely the Chern ${\cal C}=\sum_{\eta,s} {\cal C}_s^{\eta}$, spin Chern  ${\cal C}_s=\sum_{\eta,s} s~ {\cal C}_s^{\eta}/2$, valley Chern  ${\cal C}_{\eta}=\sum_{\eta,s} \eta~{\cal C}_s^{\eta}$, and spin-valley Chern  $C_{s \eta}=\sum_{\eta,s} \eta s~{\cal C}_s^{\eta}/2$ numbers.  Here ${\cal C}_s^{\eta}=\eta~ \text{sign}[\Delta_s^{\eta}]/2$ is the Pontryagin number \cite{TI1,TI2}, a topological quantity that counts how many times the vector ${\bf d}^{\eta}_s/|{\bf d}^{\eta}_s|$ wraps a sphere in momentum space (Fig. \ref{Fig1}c,d).  By varying the external parameters  $\lambda_E, \ \lambda_L$, and $\lambda_M$ the monolayer can be driven through various phase transitions as depicted in Fig. \ref{Fig1}b, where we show the topological phases in the planes $\lambda_E =0$, $\lambda_L =0$, and $\lambda_M =0$.  The electronic states quantum spin Hall insulator (QSHI), anomalous quantum Hall insulator (AQHI), band insulator (BI), antiferromagnetic insulator (AFI), and  polarized- spin quantum Hall insulator (PS-QHI) have non-zero mass gaps for all four Dirac cones. The system undergoes a topological phase transition when the band gap of at least one Dirac cone changes sign. Hence, the boundaries between insulating states are determined by the condition $\Delta^\eta_s = 0$, which defines the so called single Dirac cone phases (diagonal solid lines in Fig. \ref{Fig1}b). At the intersection between boundary lines there are two Dirac gaps closed, and the system is either in a spin, valley, or spin-valley polarized semimetal state. Finally, note that there are four points in the 3D phase diagram,  $\{\lambda_E/\lambda_{SO},\lambda_L/\lambda_{SO},\lambda_M/\lambda_{SO}\} = \{-1,-1,1\},\ \{1,1,1\},\ \{1,-1,-1\},\ \{-1,1,-1\}$, where three Dirac cones simultaneously close, corresponding to unique electronic states that remain unexplored to date.
\\

\hspace{-9pt}{\bf Nonlinear light-matter interactions in the GFM  -} 
The nonlinear dynamics of charge carriers in a given Dirac cone interacting with a linearly polarized electromagnetic plane wave impinging normally on the monolayer can be described via the density matrix ${\hat \rho}(t)$, which satisfies the equation of motion 
%
$i\hbar\ \partial \hat{\rho}(t)/\partial t= \left[{\hat H}_s^{\eta}+ \hat{H}_{i}(t),\hat{\rho}(t)\right]-\Gamma({\hat \rho(t)}-{\hat \rho}^{(0)}).$
%
Here, $\hat{H}_{i}(t)=e {\hat {\bm p}}\cdot{{\bm A}}(t)/mc$ is the field-monolayer interaction Hamiltonian in the velocity gauge \cite{Sipevelocity}, where ${\bm A}(t) = A(t)\hat{\textbf{x}}$ is the electromagnetic vector potential.  We assume that at  large times the system relaxes to the equilibrium density matrix $\hat{\rho}^{(0)}=\sum_{l,{\bm k}} f_{l {\bm k}} |l{\bm k}\rangle\langle l{\bm k} |$ with a phenomenological decay rate  $\Gamma$, where $f_{l{\bm k}}$ is the Fermi-Dirac distribution for fermions in the valence or conduction  bands ($l=\pm1$) with momentum $\hbar \textbf{k}$, and $|l{\bm k}\rangle$ are the corresponding eigenstates of the unperturbed Hamiltonian $H^{\eta}_s$ (see Methods). In general, the equation of motion cannot be solved analytically  due to the time dependence of the interaction Hamiltonian, and numerical methods are often employed. Here, instead, we use perturbation theory by expanding the density matrix ${\hat \rho}(t)=\sum_n {\hat \rho}^{(n)}(t)$ in powers of the vector potential amplitude and iteratively solve for $\rho^{(n)}(t) \propto A(t)^n$. We use the standard expression for total 
current ${\bm j}(t)=-e{\text Tr}[{\hat \rho}(t)({\hat {\bm v}}+\frac{e}{mc} {\hat {\bm A}}(t))]$, which contains contributions arising from all orders in the perturbative expansion\cite{Pedersen2017, Aversa1995}. Here, $\hat{v}=\hbar^{-1}\nabla_{{\bf k}} \hat{H}_s^{\eta}$ is the velocity operator. The $n$-th order ac electric current is given by
\begin{equation}
\label{Nthorder}
{\bm j}^{(n)}(t)=-e\textrm{Tr}[{\hat \rho}^{(n)}(t){\hat {\bm v}}]-\frac{e^2}{mc}\textrm{Tr}[{\hat \rho}^{(n-1)}(t){\bm A}(t)].
\end{equation}
Although the last term in the previous equation does not vanish in general, one can demonstrate that it is exactly canceled by another contribution arising from the first term and, therefore, its contributions do not appear in the final expressions for the conductivities. The cross-canceling of contributions involving the first and second terms in Eq. \eqref{Nthorder} is actually critical to reobtain the well know expression for the conductivity in the linear regime, given by Kubo's formalism (Methods). The
$n$-th order optical conductivity is a tensor of rank $n+1$, accounts for both intra and interband transition contributions as well as topological effects emerging from the Berry connection \cite{EzawaJapanese}, and can be directly computed after Fourier transforming the above equation and expressing the current as a product between the electric field and conductivity. 
\begin{figure}
\includegraphics[width=1.0\linewidth]{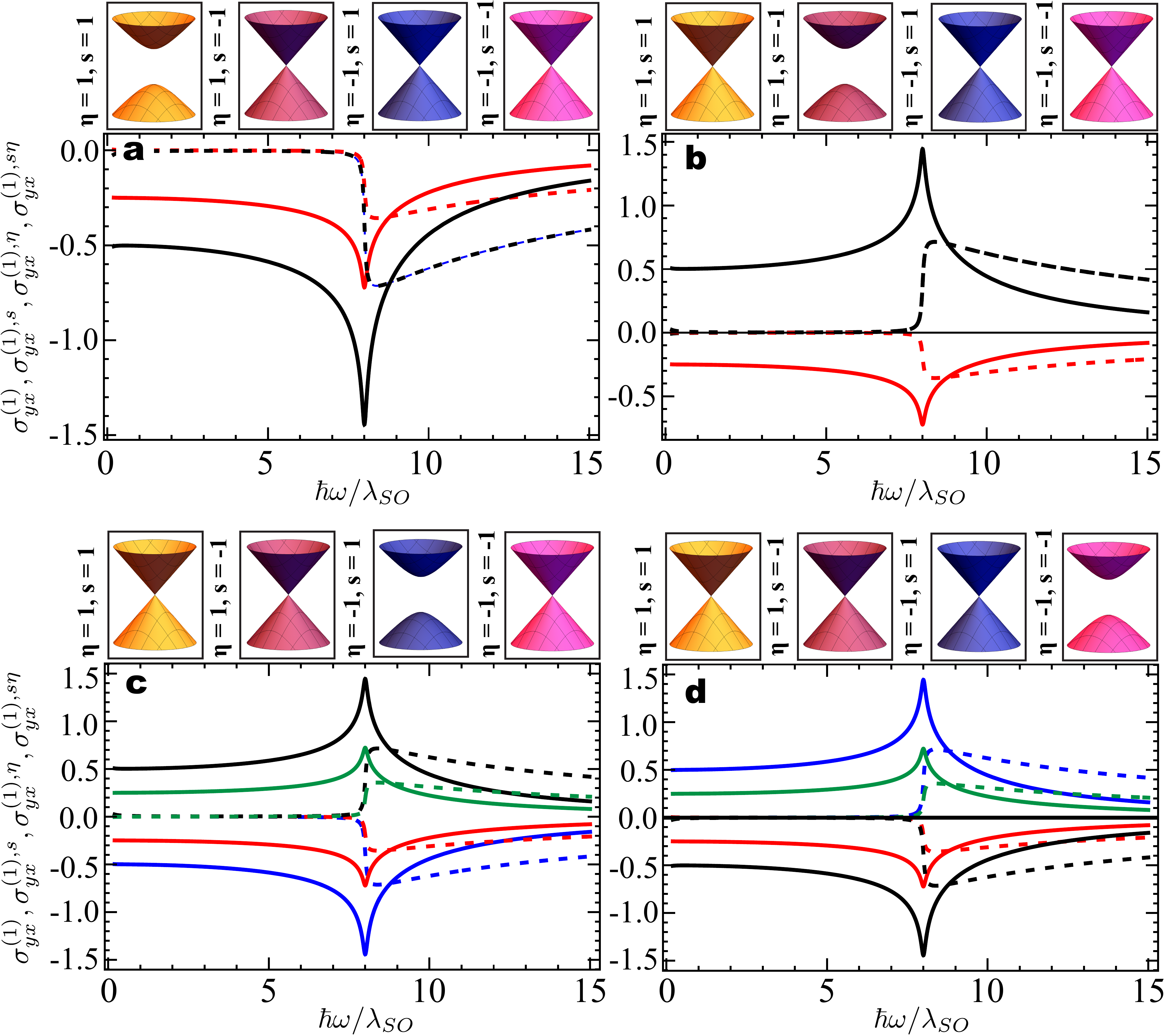}
\caption{ {\bf Identifying topological properties of individual Dirac cones in the linear regime.} Real (solid) and imaginary (dashed) components of the charge $\sigma_{yx}^{(1)}$ (blue), spin $\sigma_{yx}^{(1), s}$ (red), valley $\sigma_{yx}^{(1), \eta}$ (black), and spin-valley $\sigma_{yx}^{(1), s\eta}$ (green) transverse linear conductivities are shown for the four points in phase space where only when Dirac cone is open, namely $\{\lambda_E/\lambda_{SO},\lambda_L/\lambda_{SO},\lambda_M/\lambda_{SO}\}$ equal to (a) $\{-1,-1,1\}$ (b) $\{1,1,1 \}$ (c) $\{ 1,-1,-1\}$ (d) $\{-1,1,-1\}$. The insets at the top of each panel show the corresponding structure of Dirac cones. In panel (a) the charge and spin conductivities are identical and in panel (b) the same occurs for the valley and spin-valley conductivities. All conductivities  are expressed in units of $2\sigma_{0}/\pi$, where $\sigma_0=\alpha c/4$ and $\alpha$ is the fine-structure constant, we used $\hbar\Gamma=0.05 \lambda_{SO}$, and assumed the monolayer to be neutral.}
\label{Fig2}
\end{figure}

 The longitudinal $\sigma_{xx}^{(1)\eta, s}(\omega)$ and transverse $\sigma_{yx}^{(1)\eta, s}(\omega)$ linear optical conductivities per Dirac cone obtained from Eq. (\ref{Nthorder}) are in full accordance with previous studies, and have been investigated in details in the $\lambda_M = 0$ plane in Refs.  \citenum{Linear1}, \citenum{Tunablebandgap1}, \citenum{WiltonPRL}, \citenum{Wiltonnature}. For completeness, we explore in Fig. \ref{Fig2} the charge, spin, valley, and spin-valley transverse linear conductivities (summed over valley and spin indexes as shown in the Methods) in the four phases with only one Dirac cone open, since these have been previously overlooked in the literature.  By measuring the dynamic linear conductivities at these unique points in phase space one can investigate  topological properties emerging from each Dirac cone individually. 
Indeed, the  nonzero Dirac gap in each of the cases considered leads to Chern numbers  ${\cal C}$, ${\cal C}_{\eta}$ $=\pm 1/2$, and ${\cal C}_s$, ${\cal C}_{s\eta}$ $=\pm 1/4$, which clearly manifest in the zero frequency limit of the transverse conductivities for neutral monolayers, as shown in Fig. \ref{Fig2}. Note that in panels \ref{Fig2}a,b the charge (spin) and the valley (spin-valley) currents are identical, which implies that the nonzero Dirac gap belongs to the $K$ valley, while in panels \ref{Fig2}c,d the charge (spin) and valley (spin-valley) conductivities are the negative of each other, hence the nonzero Dirac gap belongs to $K'$ valley. Finally, the sign of the spin or spin-valley linear conductivities in the quasi-static regime allows to identify the spin of the charge carriers.
 
 The second order response of the system described by the Hamiltonian in Eq. \eqref{Hamiltonian} vanishes due to the centrosymmetry of the GFM. The first nonlinear contribution comes from the third-order response and can be described by a rank four conductivity tensor $\tilde{\sigma}_{\alpha\beta\gamma\delta}^{(3)\eta,s}(\omega_1,\omega_2,\omega_3)$, which is invariant under simultaneous permutations of the indexes $\beta,\gamma,\delta$ and frequencies $\omega_1,\omega_2,\omega_3$, where $\alpha, \beta,\gamma,\delta = x,y$ (see Methods). 
For simplicity we will concentrate in cases where $\beta = \gamma = \delta = x$ and we will assume that the incident electromagnetic radiation  is monochromatic, {\it i.e.}, ${\bm A}(t)=A_0 \cos(\omega t) ~\hat{\textbf{x}}$,  allowing us to separated the third-order conductivity into terms oscillating with frequencies $\omega$ and $3\omega$. The first term adds to the linear conductivity response and gives a correction to the fundamental harmonic  that is quadratic in the vector potential amplitude (Kerr effect), while the second one describes the nonlinear process of third harmonic generation. Therefore, up to third-order in perturbation theory, the optoelectronic conductivity of the system at the fundamental and third-harmonics due to each Dirac mass gap can be cast as 
\begin{eqnarray}
\label{Dynamic1}
\tilde{\sigma}_{\alpha x}^{\eta,s}(\omega, I_0)&=&\sigma_{\alpha x}^{(1)\eta,s}(\omega)+\frac{6\pi}{c}I_0\sigma_{\alpha x}^{(3)\eta,s}(\omega),\nonumber\\
\tilde{\sigma}_{\alpha x}^{\eta,s}(3\omega, I_0)&=&\frac{2\pi}{c}I_0\sigma_{\alpha x}^{(3)\eta,s}(3\omega)\, , 
\end{eqnarray}
where $I_{0}=\omega^2A_0^2/8\pi c^2$ is the incident field intensity and  $\tilde{\sigma}_{\alpha x}^{(3)\eta,s}(\omega)\!\!=\!\!\tilde{\sigma}_{\alpha xxx}^{(3)\eta,s}(\omega,\omega,-\omega)$, $\tilde{\sigma}_{\alpha x}^{(3)\eta,s}(3\omega)\!\!=\!\!\tilde{\sigma}_{\alpha xxx}^{(3)\eta,s}(\omega,\omega,\omega)$ is a  shorthand notation. In the next section we discuss topological fingerprints embedded in these nonlinear conductivities.
\\

\begin{figure}
\includegraphics[width = 1.0\linewidth]{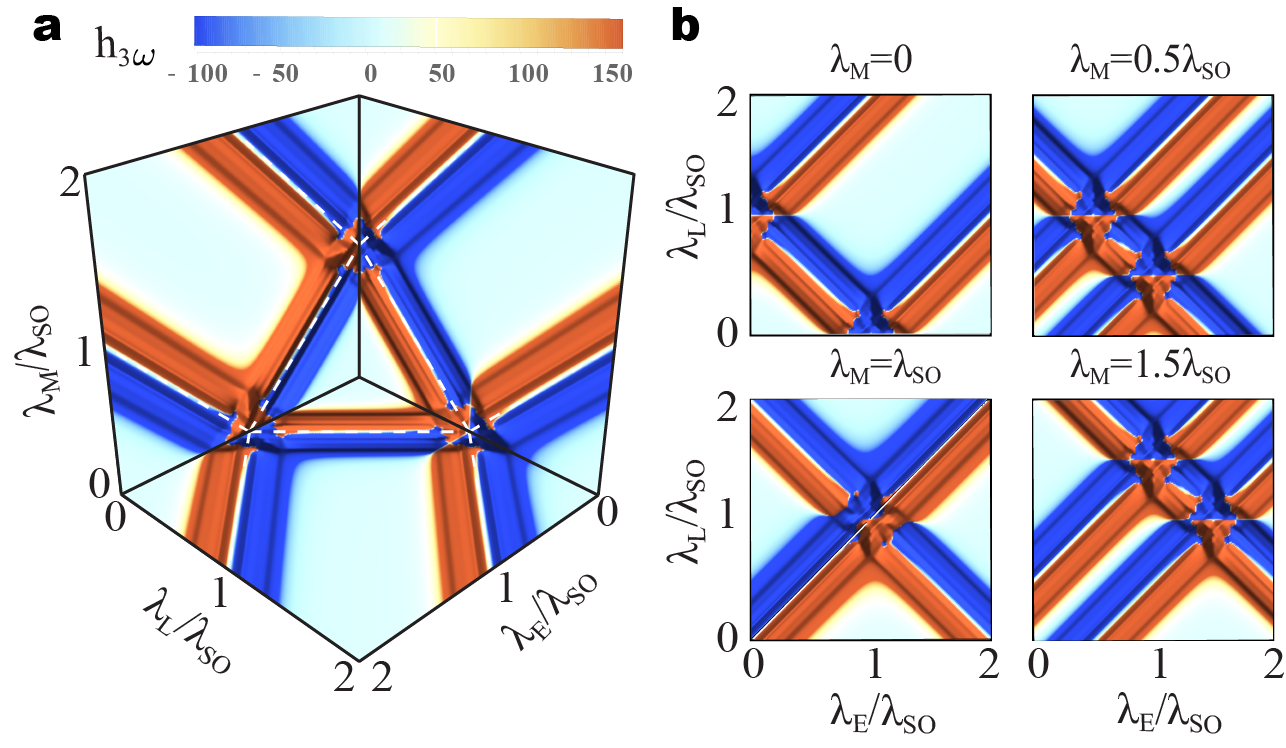}
\caption{{\bf Topological phase transitions in the third-harmonic polarization state.} (a) Phase diagram of the difference between intensities of third-harmonic emission with left and right circular polarization $h_{3\omega}$ (a.u.). (b) 2D phase diagrams of $h_{3\omega}$ at various horizontal planes with fixed value of $\lambda_M$. Here,  $\hbar \omega=0.2 \lambda_{SO}$ and $\hbar\Gamma=0.05\lambda_{SO}$.}
\label{Fig3}
\end{figure}

\hspace{-12pt}{\bf Topological signatures in the GFM nonlinear response -} 
 Let us first look into topological signatures buried in the phase of the optical field by investigating the polarization state (helicity) of light resulting from third-harmonic generation processes. We consider the low temperature regime for a neutral monolayer, {\it i.e.}, the Fermi level lies in the middle of the band gap, and assume  $\hbar\omega=0.2\lambda_{SO}$ and $\hbar \Gamma=0.05\lambda_{SO}$.  In Fig. \ref{Fig3}a we plot the phase diagram associated with the difference 
%
$h_{3\omega}=(I_L^{(3)}-I_R^{(3)})/I_0\propto 
\text{Im}\left[\sigma_{xx}^{(3)}(3\omega)^*\sigma_{yx}^{(3)}(3\omega) \right]$
%
between intensities $I_L^{(3)}$ and $I_R^{(3)}$ of light emitted with left and right circular polarization at frequency $3\omega$, respectively, where here the conductivities $\sigma_{xx}^{(3)}(3\omega)$, $\sigma_{yx}^{(3)}(3\omega)$ account for all Dirac cones (Methods).
\begin{figure*}
\includegraphics[width=0.8\linewidth]{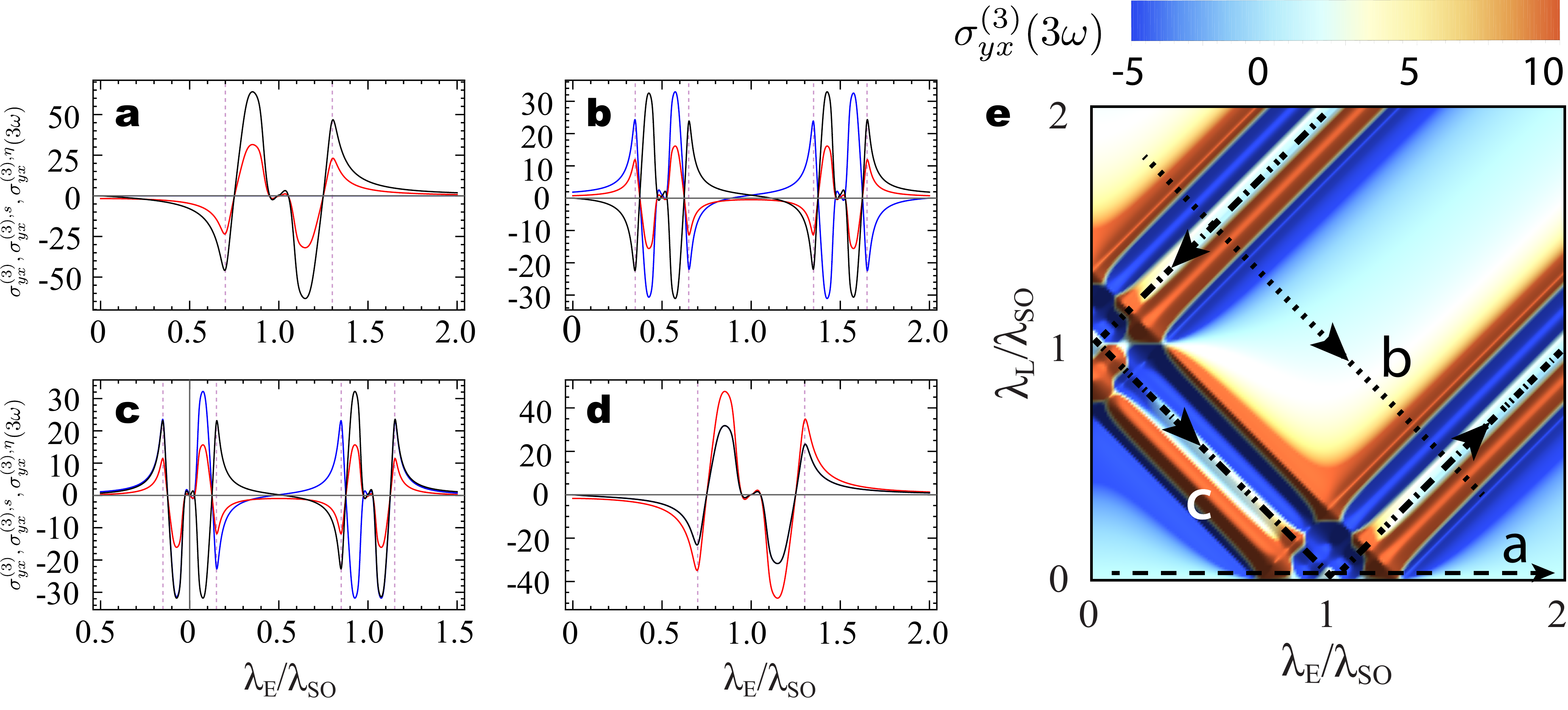}
\caption{ {\bf Nonlinear charge, spin, and valley dynamics in the graphene family materials.} Panels (a)-(d) show the third-harmonic generation for charge (blue), spin (black), and valley (red)  transverse conductivities along various topological phase transitions paths as highlighted in the (e) 2D phase diagram of $\sigma_{y x}^{(3)}(3\omega)$ for $\lambda_M=0$ plane.  While (a)-(c) paths lie on the plane $\lambda_M=0$, (d) is along the  diagonal of the 3D phase diagram connecting points ($\lambda_E/\lambda_{SO}, \lambda_L/\lambda_{SO},\lambda_M/\lambda_{SO}) = (0,0,0)\rightarrow (2,2,2)$. The paths are parametrized in terms of $\lambda_E/\lambda_{SO}$, the conductivities are expressed in the units of $\alpha^2\hbar^3c^2 v_F^2/\lambda_{SO}^4$, and the frequency and relaxation rates are the same as in Fig. \ref{Fig3}.}
\label{Fig4}
\end{figure*}
%
A direct comparison with the phase diagram in Fig. \ref{Fig1} clearly reveals the role of topology in the nonlinear optical response of the GFM.  Far from the phase boundaries the contribution from each Dirac gap to  $h_{3\omega}$ decreases as $1/|\Delta_{s}^{\eta}|^4$, resulting in weak variations of helicity in the middle of the  QSHI, BI, AQHI, PS-QHI, and AFI phases.

The effects of nonlinearity are most relevant as one approaches the semi-metallic states (white dashed lines) marking the boundary between different Chern numbers, where  $h_{3\omega}$ is dominated by the contribution from the Dirac cone with the smallest gap and the phase transition can be easily identified. These conclusions are not restricted to the planes $\lambda_E=0$,  $\lambda_L=0$, and  $\lambda_M=0$ as can be seen in Fig. \ref{Fig3}b, where we present slices of the phase diagram corresponding to fixed values of $\lambda_M$.  Note in these panels that as $\lambda_M$ increases the phase boundaries come closer to each other and at $\lambda_M/\lambda_{SO}=1$ the three phase boundaries cross at a single point, $\{\lambda_E/\lambda_{SO}, \lambda_L/\lambda_{SO}, \lambda_M/\lambda_{SO}\} = \{1 , 1, 1\}$,  corresponding to the closing of three Dirac gaps. Finally, we mention that the variations of $h_{3\omega}$ across a phase transition depend on whether the initial and final phases have trivial charge transport properties (Chern number ${\cal{C}}$ equal to zero). For instance, in the $\lambda_M=0$ we observe that $h_{3\omega}$ is a symmetric function around $\lambda_E/\lambda_{SO}=1$ when we consider the topological phase transition between the  QSHI and BI phases, both with ${\cal{C}} = 0$. On the other hand,  a remarkably asymmetric behavior takes place  around $\lambda_L/\lambda_{SO}=1$ (which is energetically equivalent to the previous case) when one considers a topological phase transition involving the QSHI and the AQHI (${\cal{C}} = -2$) phases. 

 The phase diagram of $h_{3\omega}$ includes contributions from both third-harmonic longitudinal and transverse conductivities. However, only the latter encodes topological features of the GFM and deserves to be independently studied. To this end one could, for instance, use a linear polarizer filter to block the co-polarized scattered field and detect only its cross-polarized component, which is proportional to $\sigma_{yx}^{(3)}$. In Fig. \ref{Fig4} we investigate the charge conductivity $\sigma_{yx}^{(3)}(3\omega)$  along various paths in the phase diagram (parametrized in terms of $\lambda_E/\lambda_{SO}$). For completeness we also show the associated spin  $\sigma_{yx}^{(3), \textrm{s}}(3\omega)$ and valley $\sigma_{yx}^{(3), \eta}(3\omega)$ third harmonic transverse conductivities.

  In panel \ref{Fig4}a we plot the conductivities along the $\lambda_E$ axes as we move from the QSHI to the BI phase by closing two Dirac gaps when $\lambda_E/\lambda_{SO}=1$. In this case the Chern numbers (${\cal{C}}, {\cal{C}}_s, {\cal{C}}_\eta$) change from ($0,1,0$) to ($0,0,-2$) and we observe that the charge conductivity is always zero.
The resonant peaks and dips near $\lambda_E/\lambda_{SO}=1$ in the spin and valley conductivities  divulge information about the change in  ${\cal{C}}_s$ and ${\cal{C}}_\eta$ across the phase transition. 
For instance, as the system is driven from the QSHI to the BI phase the resonance at $3\hbar\omega = 2\Delta_s^\eta$ (dashed vertical lines) in $\sigma_{yx}^{(3), \textrm{s}}(3\omega)$ and $\sigma_{yx}^{(3), \eta}(3\omega)$ flips from negative to positive values, which is a signature indicating a decrease in the corresponding Chern numbers across the phase transition. Also, the magnitude of the resonance peaks and dips in the valley conductivity is twice as those of the spin conductivity, which coincides with the fact that the change in ${\cal{C}}_{\eta}$ is two times that in ${\cal{C}}_s$. These features are general and can be applied to any paths in the phase diagram regardless of the nature of the transition involved.  

In panel \ref{Fig4}b we plot the conductivities as the system moves from the AQHI to the BI phase via the PS-QHI phase. Along this path the monolayer crosses two phase boundaries, with Chern numbers (${\cal{C}}, {\cal{C}}_s, {\cal{C}}_\eta$) changing as $(-2,0,0)$ $\rightarrow$ ($-1,1/2,-1$)$ \rightarrow$ ($0,0,-2$). Note that the signs of the resonances at $3\hbar\omega = 2\Delta_s^\eta$  for the valley conductivity are opposite to those for the charge conductivity, while their strengths are the same, meaning that the variations in ${\cal C}$ and ${\cal C}_{\eta}$ across the phase transition are equal in magnitude. The sign of the third harmonic resonance for the charge (valley) conductivity changes from negative (positive) in the AQHI phase to positive (negative) in the PS-QHI phase, which agrees with an increase (decrease) in ${\cal{C}}$ (${\cal{C}}_\eta$). The same feature repeats near the phase transition between the PS-QHI and BI phases, signaling another increase (decrease) in ${\cal{C}}$ (${\cal{C}}_\eta$). Moreover, the behavior of the spin conductivity in Fig. \ref{Fig4}b reflects the fact that ${\cal C}_{s}$ increases along the first phase boundary and then decreases along the second transition, further confirming that the nonlinear resonant behavior of the conductivities encode topology fingerprints. 

In the third panel of Fig. \ref{Fig4} the system is driven along the phase boundaries highlighted in path (c), where at least one of the Dirac gaps is closed all the time and the Chern numbers change according to $(-3/2,1/4,-1/2)$ $\rightarrow$ ($-1/2,3/4,-1/2$) $ \rightarrow$ ($-1/2,1/4,-3/2$). It is important to notice that ${\cal C}$ does not change near $\lambda_E/\lambda_{SO}=1$ and ${\cal C}_{\eta}$ remains the same across the transition at $\lambda_{L}/\lambda_{SO}=1$ ($\lambda_E = 0$ in the parametrized curve).  This fact is clearly reflected in Fig. \ref{Fig4}c, where the third harmonic resonances associated with  charge (valley) conductivity have the same signs as the system is driven through $\lambda_E/\lambda_{SO}=1$ ($\lambda_L/\lambda_{SO}=1$). Finally, in Fig. \ref{Fig4}d we plot the conductivities for a path (not shown in Fig. \ref{Fig4}e)  going from the origin of the phase diagram to the point $\{\lambda_E/\lambda_{SO}, \lambda_L/\lambda_{SO}, \lambda_M/\lambda_{SO}\} = \{2 , 2, 2\}$ across the diagonal $\lambda_E = \lambda_L  = \lambda_M$. Once again, the  behavior of the conductivities   and the change in Chern numbers agrees well with signatures we explained above.

\begin{figure}
\includegraphics[width=1\linewidth]{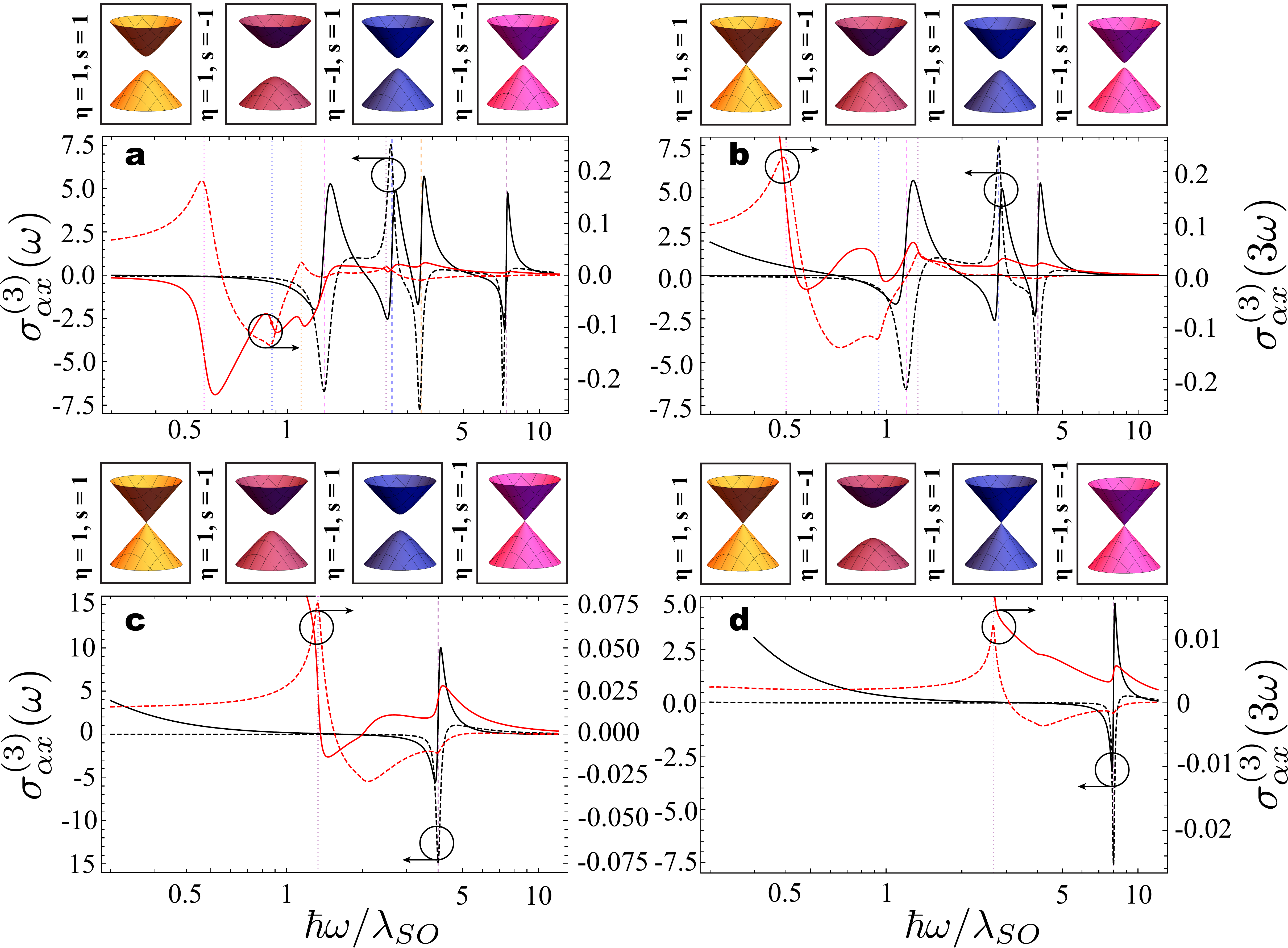}
\caption{ {\bf Nonlinear spectral response of spin-orbit coupled graphene family materials.} Real part of the third-order optical conductivities corresponding to Kerr effect (black, frequency $\omega$) and third harmonic generation (red, frequency $3\omega$) in both longitudinal (solid, $\alpha = x$) and and transverse (dashed, $\alpha = y$) directions for $\{\lambda_E/\lambda_{SO},\lambda_L/\lambda_{SO},\lambda_M/\lambda_{SO}\}$ equal to (a) $\{1.5,1.2,0\}$, (b) $\{0.7,0.3,0\}$, (c) $\{0,1,0\} $, and (d) $\{1,1,1\}$.  The dashed (dotted) vertical gridlines indicate the resonance conditions for the third (fundamental) harmonic.  Conductivities are expressed in units of $\alpha^2 \hbar^5 \omega^2c^2 v_F^2/\lambda_{SO}^6 $ and we assume $\hbar \Gamma/\lambda_{SO}=0.1$.}
\label{Fig5}
\end{figure}

Next we show how one can identify  signatures of the Chern numbers  in each topological phase by investigating the nonlinear spectral response of the GFM. In Fig. \ref{Fig5} we present the frequency dispersion of the real part of the third-order optoelectronic longitudinal and transverse conductivities for both fundamental and third harmonics at four different points in phase space. These correspond to cases where all gaps are open (Fig. \ref{Fig5}a) or one (Fig. \ref{Fig5}b),  two (Fig. \ref{Fig5}c), or three (Fig. \ref{Fig5}d) Dirac gaps are closed.  The corresponding curves for the imaginary part (not shown) of the conductivities can be obtained directly from our formalism or via generalized Kramers-Kronig relations for nonlinear systems \cite{Hutchings1992}. We expect that the nonlinear conductivities should have non-negligible contributions when at least one Dirac gap matches one or three times the energy of the incident photons. Indeed,  Fig. \ref{Fig5} shows that Kerr $\sigma_{\alpha x}^{(3)}(\omega)$ and third harmonic $\sigma_{\alpha x}^{(3)}(3\omega) $ conductivities present localized spectral resonances at $2\Delta_s^{\eta} =\hbar \omega$  (dashed vertical gridlines) and $2\Delta_s^{\eta} =3\hbar \omega$ (dotted vertical gridlines), respectively, with richer and more involved spectral responses in the cases where more Dirac cones are open. We mention that  although the third-harmonic generation conductivity also includes resonances near $\omega$ and $2\omega$, they are difficult to identify in the spectrum because their magnitude are significantly smaller than the resonances near $3\omega$. 

Note that near resonances the longitudinal Kerr and third harmonic conductivities always shows normal and anomalous dispersion, respectively, regardless of the resonant Dirac cone. This is due to the fact that both of these conductivities depend only on the absolute value of the resonant mass gap, hence not allowing to directly distinguish between energetically equivalent but topologically different points in the phase space. On the other hand, the transverse Hall conductivity is sensitive to the sign of the resonant $\Delta_{s}^\eta$ and, therefore, encodes information of the corresponding Pontryagin number ${\cal C}_s^{\eta}$.  Note in Fig. \ref{Fig5}, for instance,   that $\sigma_{yx}^{(3)}(\omega)$ and $\sigma_{yx}^{(3)}(3\omega)$  have positive and negative resonances depending on the resonant Dirac gap. Therefore, just as the linear Hall conductivity allows to probe topological properties of the GFM at finite frequencies \cite{WiltonPRB}, so it does at the nonlinear regime. Indeed, the Chern number associated to the open Dirac cones at any point in phase space can be computed by summing the sign of $\sigma_{yx}^{(3)}(\omega)$  or $\sigma_{yx}^{(3)}(3\omega)$ precisely at the resonant frequencies, accounting appropriately for degeneracy among mass gaps, and multiplying the result by $+1/2$ or $-1/2$, respectively. Consider, for example, the case in Fig. \ref{Fig5}a. Note that in this case the third-order conductivity at the fundamental (third) harmonic evaluates to negative (positive) values in three of the four resonances. Thus, the sum of their signs equals $-2$ ($+2$) and results in a Chern number ${\cal C}=-1$, as expected for the PS-QHI phase in the $\lambda_M = 0$ plane. Note that this argument can be extended to the nonlinear spin, valley, and spin-valley Hall currents, allowing for obtaining the full set of Chern numbers characterizing the topological phases in buckled two-dimensional semiconductors. In addition, unlike the linear case where to obtain the Chern number one needs to measure the linear Hall conductivity at various points around a resonance to obtain the sign of its derivative  \cite{WiltonPRB}, here we need to evaluate $\sigma_{yx}^{(3)}(3\omega)$ at most at the four (in the non-degenerate case) resonant frequencies. Finally, we mention that the dependence of the transverse (longitudinal) conductivities on $\textrm{sign}[\Delta_{s}^\eta]$  ($|\Delta_{s}^\eta|$) are not exclusive to the third order nonlinear case and similar conclusions should  hold for higher order contributions as well.%
\\

\hspace{-12pt} {\bf Conclusions}

 In summary, we developed a comprehensive investigation of nonlinear light-matter interactions in spin-orbit coupled graphene materials, with focus on third harmonic generation and nonlinear corrections to the fundamental harmonic (Kerr effect). We have shown that by controlling various external parameters the monolayer can be driven through different phase transitions to explore the interplay between nonlinear effects and topological properties of two-dimensional materials. Specifically, we demonstrated that the helicity of the nonlinear scattered field encodes information about the location of the phase transitions boundaries, the dependence of the cross-polarized component of the field (proportional to the transverse conductivity) on the external parameters near resonances divulges details about the topological phases involved in the transition, and the dispersive nonlinear response of the system enables characterization of the Chern number in each phase. The recent progress in synthesis of topological semiconductors of the graphene family materials \cite{Silicene, Germanene, Stanene, Plumbene} and well established nonlinear characterization photonic techniques \cite{Kerr, Gnonlinear2,Exp2,Exp3,Exp4,Exp5} suggest that our findings can be  accessed experimentally with current technologies. We envision that the the effects predicted here will greatly impact research at the crossroads  between nonlinear optics, topological materials, spintronics, and valleytronics.
\\

\hspace{-12pt} {\bf Methods}

 In order to compute the nonlinear conductivity we need the unperturbed eigenvectors and the velocity matrix elements associated to the Hamiltonian in Eq. \eqref{Hamiltonian}.
In the following we will omit the valley and spin indexes $\eta$ and $s$ whenever possible, and the reader should keep in mind that, unless otherwise stated, all results presented are valid for a single Dirac cone. The total conductivities are given by summing the contributions of all mass gaps in the band structure. The eigenvector associated to the Hamiltonian in Eq. \eqref{Hamiltonian} has the form
\begin{equation}
\label{Eigen}
|l \bm{k}\rangle=\frac{\hbar v_F |{\bf k}|}{\left[2\epsilon\left(\epsilon+l \Delta\right)\right]^{1/2}}\left(\begin{matrix}
\frac{\epsilon+l \Delta}{\eta \hbar v_F |{\bf k}| }e^{-i \eta \theta}\\ l
\end{matrix}\right),  
\end{equation}
where $|{\bf k}| = \sqrt{k_x^2+k_y^2}$, $\epsilon_l = l\epsilon=l\sqrt{\hbar^2v_F^2 |{\bf k}|^2+\Delta^2}$,  $l$ is the conduction ($l=+1$) and valence ($l=-1$) band index, and $\theta=\tan^{-1} (k_y/k_x)$.  The matrix elements $\langle l {\bf k} |\hat{v}_{\alpha}| l'{\bf k}\rangle$ of the velocity operator $\hat{v}_{\alpha}= \hbar^{-1} \partial \hat{H}^\eta_s/\partial k_\alpha$ are given by
\begin{eqnarray}
\label{Matrixelementx}
\langle 1\bm{k}|\hat{v}_{x}|-1\bm{k}\rangle&=&\langle -1\bm{k}|\hat{v}_{x}|1\bm{k}\rangle^*=-v_F\frac{\Delta \cos \theta +i\epsilon \sin \theta}{\epsilon},\cr
\langle1\bm{k}|\hat{v}_{x}|1\bm{k}\rangle&=&-\langle-1\bm{k}|\hat{v}_{x}|-1\bm{k}\rangle=\frac{\hbar v_F^2 k\cos\theta}{\epsilon},\cr
\langle 1\bm{k}|\hat{v}_{y}|-1\bm{k}\rangle&=&\langle -1\bm{k}|\hat{v}_{y}|1\bm{k}\rangle^*=-v_F\frac{i\epsilon \cos \theta -\Delta \sin \theta}{\epsilon},\cr
\langle1\bm{k}|\hat{v}_{y}|1\bm{k}\rangle&=&-\langle-1\bm{k}|\hat{v}_{y}|-1\bm{k}\rangle=\frac{\hbar v_F^2 k \sin\theta}{\epsilon}.
\end{eqnarray}

We give a brief derivation for both the linear and third order optical conductivity terms for a particular Dirac cone.  To this end,  the equation of motion $i\hbar\partial \hat{\rho}(t)/\partial t= \left[{\hat H}_s^{\eta}+ \hat{H}_{i}(t),\hat{\rho}(t)\right]$ for the density matrix operator  is solved perturbatively by expanding the density matrix in powers of the potential vector amplitude ${\hat \rho}(t)=\sum_n {\hat \rho}^{(n)}(t)$ with ${\hat \rho}^{(n)}(t) \propto A^n$. To leading order in the potential vector, the first order density matrix upon Fourier transformation to the frequency domain can be cast as 
\begin{equation}
\label{rho1}
\langle l{\bf k}|\rho^{(1)}(\omega)|l'{\bf k}\rangle=\frac{-ie E_{\alpha}(\omega)}{\omega}\frac{f_{l'{\bf k}}-f_{l{\bf k}}}{\epsilon_{l'}-\epsilon_{l}+\hbar\omega}\langle l{\bf k}|\hat{v}_{\alpha}|l'{\bf k}\rangle, 
\end{equation}
where $E_{\alpha}(\omega) = i\omega A_\alpha(\omega)/c$. Substituting $\hat{\rho}^{(1)}(\omega)$ in equation \eqref{Nthorder} and making use of the identity $[\hbar\omega\left(\epsilon_{l'}-\epsilon_{l}+\hbar\omega\right)]^{-1}=[\epsilon_{l'}-\epsilon_{l}]^{-1}[(\hbar\omega)^{-1}-(\epsilon_{l'}-\epsilon_{l}+\hbar\omega)^{-1}]$, we eliminate the first term in the current $\propto Tr(\rho^{(0)})$ and re-obtain the well know Kubo formula after expressing the current as the product of conductivity and electric field
\begin{equation}
\label{linear}
\sigma_{\alpha \beta}^{(1), \eta, s}(\omega)=-4i \hbar^2 \sigma_0\sum\limits_{{\bf k},l,l'}\frac{f_{l'{\bf k}}-f_{l{\bf k}}}{\epsilon_{l'}-\epsilon_{l}}\frac{\langle l{\bf k}|\hat{v}_{\beta}|l'{\bf k}\rangle\langle l'{\bf k}|\hat{v}_{\alpha}|l{\bf k}\rangle}{\epsilon_{l'}-\epsilon_{l}+\hbar\omega}, 
\end{equation}
where $\sigma_0 = \alpha c/4$ and the phenomenological relaxation rate is accounted for by replacing $\omega \rightarrow \omega +i\Gamma$. Equation (\ref{linear}) accounts for both interband and intraband transitions. The interband contribution is straightforward and follows directly from (\ref{linear}) by setting $l'\neq l $. The intraband contribution can be derived by equating enforcing $l'=l$, in which case $(f_{l'}-f_{l})/(\epsilon_{l'}-\epsilon_l) \rightarrow \partial f_{lk}/\partial \epsilon_l$. The total charge, spin, and valley linear conductivities are then computed as  $\sigma_{\alpha x}^{(1)}(\omega) = \sum_{\eta,s} \sigma_{\alpha x}^{(1)\eta, s}(\omega)$, $\sigma_{\alpha x}^{(1), \textrm{s}}(\omega) = \sum_{\eta,s} s\sigma_{\alpha x}^{(1)\eta, s}(\omega)/2$,  $\sigma_{\alpha x}^{(1), \eta}(\omega) = \sum_{\eta,s} \eta\sigma_{\alpha x}^{(1)\eta, s}(\omega)$, and $\sigma_{\alpha x}^{(1), s\eta}(\omega) = \sum_{\eta,s} s\eta\sigma_{\alpha x}^{(1)\eta, s}(\omega)/2$, respectively. 

For the third order conductivities we first solve the equation of motion for third order correction $\rho^{(3)}(t)$, which is expressed in the Fourier space as
\begin{widetext}
\begin{eqnarray}
\langle l{\bf k}|\rho^{(3)}(\omega)|l'{\bf k}\rangle\hspace{-1mm}= \hspace{-1mm}\frac{e^3 E_{\alpha}(\omega_1)E_{\beta}(\omega_2)E_{\gamma}(\omega_3)\!/\omega_1\omega_2\omega_3}{i\left(\epsilon_{l'}-\epsilon_{l}+\hbar\omega_1+\hbar\omega_2+\hbar\omega_3\right)}\!\!\!
\sum\limits_{l'',l'''}
&&\!\!\frac{\langle l{\bf k}|\hat{v}_{\delta}\!|l'''{\bf k}\rangle\!\langle l'''{\bf k}|\hat{v}_{\gamma}\!|l''{\bf k}\rangle\!\langle l''{\bf k}|\hat{v}_{\beta}\!|l'{\bf k}\rangle}{\epsilon_{l'}-\epsilon_{l'''}+\hbar\omega_1+\hbar\omega_2}\!\!  \left(\!\! \frac{f_{l'{\bf k}}-f_{l''{\bf k}}}{\epsilon_{l'}\!-\!\epsilon_{l''}\!+\!\hbar\omega_1\!}\!-\!\frac{f_{l''{\bf k}}-f_{l'''{\bf k}}}{\epsilon_{l''}\!-\!\epsilon_{l'''}\!+\!\hbar\omega_2}\! \right) \cr
&&\!\!\!\!\!\!\!\!-
\frac{\langle l{\bf k}|\hat{v}_{\delta}\!|l'''{\bf k}\rangle\!\langle l'''{\bf k}|\hat{v}_{\gamma}\!|l''{\bf k}\rangle\!\langle l''{\bf k}|\hat{v}_{\beta}\!|l'{\bf k}\rangle}{\epsilon_{l''}-\epsilon_{l}+\hbar\omega_2+\hbar\omega_3}\!\!\left(\!\! \frac{f_{l''{\bf k}}-f_{l'''{\bf k}}}{\epsilon_{l''}\!-\!\epsilon_{l'''}\!+\!\hbar\omega_2}\!-\!\frac{f_{l'''{\bf k}}-f_{l{\bf k}}}{\epsilon_{l'''}\!-\!\epsilon_{l}\!+\!\hbar\omega_3}\! \right).\nonumber\\
\end{eqnarray}
\end{widetext}

Substituting the above expression in Eq. \eqref{Nthorder} and using the same identity as before with the replacement $\omega \rightarrow \omega_1+\omega_2+\omega_3$, we obtain the third order optical conductivity per Dirac cone as
%
\begin{eqnarray}
\label{third}
\!\!\!\!\!\!\!\!\sigma_{\alpha\beta\gamma\delta}^{(3)}(\omega_1,\omega_2,\omega_3)&=&\dfrac{-e \hbar(\omega_1+\omega_2+\omega_3)}{ E_{\alpha}(\omega_1)E_{\beta}(\omega_2)E_{\gamma}(\omega_3)} \times \cr
&\times&
 \sum\limits_{{\bf k}, l, l'} \frac{\langle l{\bf k}|\hat{v}_{\alpha}|l'{\bf k}\rangle}{\epsilon_{l'}-\epsilon_{l}} \langle l'{\bf k}|\rho^{(3)}(\omega)|l{\bf k}\rangle. 
\end{eqnarray}
%
The phenomenological relaxation rate can be accounted in a similar manner as in the linear case. Note that, Eq.  (\ref{third}) includes contributions from Berry connection, as well as intra and interband transitions \cite{Sipevelocity, Aversa1995}.
The nonlinear contributions to the fundamental harmonic associated to the Kerr effect follow by setting $\omega_1 = \omega_2 = -\omega_3$, while the third harmonic conductivity is can be derived for $\omega_1 = \omega_2 = \omega_3$. Similarly to the linear case the total charge, spin, valley, and spin-valley conductivities due to all Dirac cones are given by $\sigma_{\alpha x}^{(3)}(n\omega) = \sum_{\eta,s} \sigma_{\alpha x}^{(3)\eta, s}(n\omega)$, $\sigma_{\alpha x}^{(3), \textrm{s}}(n\omega) = \sum_{\eta,s} s\sigma_{\alpha x}^{(3)\eta, s}(n\omega)/2$,  $\sigma_{\alpha x}^{(3), \eta}(n\omega) = \sum_{\eta,s} \eta\sigma_{\alpha x}^{(3)\eta, s}(n\omega)$, and $\sigma_{\alpha x}^{(3), s\eta}(n\omega) = \sum_{\eta,s} s\eta\sigma_{\alpha x}^{(3)\eta, s}(n\omega)/2$, respectively, where $n = 1, 3$.
\\

\hspace{-12pt} {\bf Data availability}

 The data that support the plots within this paper and other findings of this study are available from the corresponding author on reasonable request.

\vspace{12pt}

\hspace{-12pt} {\bf Acknowledgements}

Research  presented  in  this  article  was  supported  by  the Laboratory  Directed  Research  and  Development  program of  Los  Alamos  National  Laboratory  under  project  number 20190574ECR.  The authors also thank the Center for Nonlinear Studies at LANL for financial  support under project 20190495CR.
\\

\hspace{-12pt} {\bf Author contributions}

 W.K.-K. proposed the study and developed the theory together with R.K.M. R.K.M. performed the numerical calculations. Both authors discussed the results and contributed to the final version of the manuscript. W.K.-K. supervised the entire study.

$^*$ Corresponding author: kortkamp@lanl.gov

\end{document}